\documentclass[a4paper,10pt,aps,prl,twocolumn,superscriptaddress,nofootinbib,balancelastpage]{revtex4-1}

\usepackage{amsmath,amssymb,graphicx}
\usepackage[T1]{fontenc}
\usepackage[utf8]{inputenc}
\usepackage{color}
\usepackage[unicode=true,pdfusetitle,bookmarks=true,bookmarksnumbered=false,bookmarksopen=false,breaklinks=true,pdfborder={0 0 0},backref=false,colorlinks=true]{hyperref}
\usepackage[normalem]{ulem} 
\hypersetup{citecolor=blue,filecolor=blue,linkcolor=blue,urlcolor=blue}

\begin{document}

\title{Bias and scatter in the Hubble diagram from cosmological large-scale structure}

\author{Julian Adamek}
\email[]{julian.adamek@qmul.ac.uk}
\affiliation{School of Physics \& Astronomy, Queen Mary University of London, 327 Mile End Road, London E1 4NS, UK}

\author{Chris Clarkson}
\email[]{chris.clarkson@qmul.ac.uk}
\affiliation{School of Physics \& Astronomy, Queen Mary University of London, 327 Mile End Road, London E1 4NS, UK}
\affiliation{Department of Physics \& Astronomy, University of the Western Cape, Cape Town 7535, South Africa}
\affiliation{Department of Mathematics \& Applied Mathematics, University of Cape Town, Rondebosch 7701, South Africa}

\author{Louis Coates}
\email[]{l.j.c.coates@qmul.ac.uk}
\affiliation{School of Physics \& Astronomy, Queen Mary University of London, 327 Mile End Road, London E1 4NS, UK}

\author{Ruth Durrer}
\email[]{ruth.durrer@unige.ch}
\affiliation{D\'epartement de Physique Th\'eorique \& Center for Astroparticle Physics, Universit\'e de Gen\`eve, 24 Quai E.\ Ansermet, 1211 Gen\`eve 4, Switzerland}

\author{Martin Kunz}
\email[]{martin.kunz@unige.ch}
\affiliation{D\'epartement de Physique Th\'eorique \& Center for Astroparticle Physics, Universit\'e de Gen\`eve, 24 Quai E.\ Ansermet, 1211 Gen\`eve 4, Switzerland}

\date{\today}

\begin{abstract}

An important part of cosmological model fitting relies on correlating distance indicators of objects (for example type Ia supernovae) with their redshift, often illustrated on a Hubble diagram.
Comparing the observed correlation with a homogeneous model is one of the key pieces of evidence for dark energy. The presence of cosmic structures introduces a bias and scatter, mainly due to gravitational lensing and peculiar velocities, but also due to smaller non-linear relativistic contributions which are more difficult to account for.
For the first time we perform ray tracing onto halos in a relativistic N-body simulation. Our simulation is the largest that takes into account all leading-order corrections from general relativity in the evolution of structure, and we present a novel methodology for working out the non-linear projection of that structure
onto the observer's past light cone. We show that the mean of the bias in the Hubble diagram is indeed as small as expected from perturbation theory. However, the distribution of sources is significantly skewed with a very long tail of highly magnified objects and we illustrate that the bias of cosmological parameters strongly depends on the function of distance which we consider.

\end{abstract}

\maketitle

\section{Introduction}

How light propagation in the inhomogeneous Universe affects our interpretation of observed data in terms of a cosmic expansion history is an important problem in cosmology~\cite{Clarkson:2011br,Kaiser:2015iia,Fleury:2016fda}. Most lines of sight pass mainly through low-density regions and most sources are therefore de-magnified compared to the mean, with relatively few significantly magnified ones in compensation. Do these competing effects exactly cancel, or do they leave a residual bias in the Hubble diagram? 

The main factors are gravitational lensing of the source by the intervening matter
and the fact that redshift is affected by peculiar motion. These effects have relatively
simple Newtonian counterparts, but there are a host of complicated relativistic corrections once light propagation is worked out in more detail.
There are selection effects too: we are much more likely to observe sources in halos; some objects are obscured from view by bright clusters; and so on.  Within perturbation theory it is relatively easy to predict the expectation value of the bias in the Hubble diagram for a random direction: this is significant for the luminosity distance as a function of redshift, $D_L(z)$, but remains small for the function $1/D_L^2(z)$~\cite{Kibble:2004tm,BenDayan:2012wi,Clarkson:2014pda,Bonvin:2015uha,Bonvin:2015kea,Kaiser:2015iia}.
Depending on what is used for model fitting this alone can lead to per-cent changes in parameter estimation~\cite{Fleury:2016fda}. As we shall show, the shape of the probability distribution function (PDF) for any of those observables is much more important than the shift in their mean.  
Because the perturbative prediction of the shift of the mean in $D_L(z)$ requires a relativistic calculation, simulating the full PDF will also have important features that can only be accessed using a relativistic ray-tracer.

In this Letter we address this question for the first time using a comprehensive non-perturbative relativistic numerical calculation. In a first step, we carry out a high-resolution N-body simulation of cosmic structure formation using the relativistic particle-mesh N-body code \textit{gevolution} \cite{Adamek:2015eda}. For a chosen observer, we extract the complete particle and metric information on the past light cone.
A halo catalog is constructed from the particle distribution, and we use a ray-tracing algorithm to characterize the null rays that connect each source with the observer non-perturbatively using the metric on the light cone. This provides us with a theoretical model of the observed Hubble diagram and the PDF for observables such as distance or apparent magnitude.

Light-cone effects have been studied with Newtonian simulations (e.g.\ \cite{Jain:1999ir,Teyssier:2008zd,Fosalba:2013mra,Borzyszkowski:2017ayl,Breton:2018wzk}),
using the Newtonian limit of general relativity (GR) and often many additional approximations.
The former is expected to be very accurate in the context of $\Lambda$CDM cosmology, but certain gauge issues on very large scales are commonly ignored \cite{Fidler:2017pnb,Adamek:2017kir}.
Ray tracing has also been applied in post-Newtonian settings \cite{Sanghai:2017yyn} and in the context of numerical relativity \cite{Giblin:2016mjp}. While these approaches require fewer assumptions, the cosmological scenarios studied so far remain crude. The scope of our work considerably extends beyond each of these previous achievements. Our relativistic simulations model spacetime and the matter distribution with unprecedented accuracy, and our novel ray-tracing method maps these to a statistical ensemble of observed sources using
equations that are formally exact in the scalar sector of gravity and that additionally include frame-dragging.
For simplicity we are considering dark matter halos as a proxy for the astrophysical objects that can be observed and avoid the complications of baryonic effects.

\section{Simulations}

Our results are based on a relativistic N-body simulation for a cosmological volume of $(2.4~\mathrm{Gpc}/h)^3$ containing an unprecedented $4.5 \times 10^{11}$ mass elements of $2.6 \times 10^9~M_\odot/h$. The metric is sampled on a regular Cartesian grid of $7680^3$ points, providing a spatial resolution of $312.5~\mathrm{kpc}/h$. This allows a robust detection of dark matter halos down to about $5 \times 10^{11}~M_\odot/h$. We choose a baseline $\Lambda$CDM cosmology with $h = 0.67556$, $\Omega_c = 0.2638$, $\Omega_b = 0.048275$, and a 
radiation density that includes massless neutrinos with $N_\mathrm{eff} = 3.046$. Linear initial conditions are computed with \textit{CLASS} \cite{Blas:2011rf} at redshift $z_\mathrm{ini} = 127$, assuming a primordial power spectrum with amplitude $A_s = 2.215 \times 10^{-9}$ (at the pivot scale $0.05~\mathrm{Mpc}^{-1}$) and spectral index $n_s = 0.9619$.

In order to extract observations we choose an arbitrary observer position and identify a circular pencil beam covering $450$ deg$^2$ of the past light cone to a comoving distance of $4.5~\mathrm{Gpc}/h$.
This is achieved by allowing the pencil beam -- which occupies only one third of the available volume -- to cross through the (periodic) simulation box twice while carefully avoiding replications. Finite-volume effects are detemined by the scale of the box and can only affect sources beyond redshift $z \sim 1$.
As the simulation proceeds, we record the position and velocity of each particle as it crosses the light cone of the background model -- sufficient information to construct a linear segment of the world line that crosses the \textit{true} light cone.

Metric perturbations are instead recorded on spherical shells of fixed comoving radius, spaced at the grid resolution of $312.5~\mathrm{kpc}/h$ and discretized using  HEALPix \cite{Gorski:2004by}, at a spatial resolution comparable to the Cartesian mesh.
We identify the three simulation time steps for which the coordinate times are closest to the look-back time, and record the respective values of the metric perturbations interpolated to the pixel locations (using cloud-in-cell interpolation).
Since the Shapiro delay is never larger than our time step we have the full knowledge of the metric on the \textit{true} light cone, at the full resolution (in space and time) of our simulation.

In a first post-processing step we use the \textit{ROCKSTAR} friends-of-friends halo finder \cite{Behroozi:2011ju} to construct a halo catalog on the light cone. The $450$ deg$^2$ field of view contains about $11$ million halos above our mass cut of $5 \times 10^{11}~M_\odot/h$. We then determine the observed redshift and luminosity distance for each halo by numerically integrating the geodesic and Sachs equations
for the ray that connects the halo with the observation event. Similar to \cite{Breton:2018wzk} we do not resort to the Born approximation and instead employ a shooting method to determine the correct boundary conditions for each ray. The method starts with the zeroth-order guess for the observed angle, integrates the ray from the observer backwards in time until it reaches the comoving distance of the halo, and 
corrects the shooting angle according to the distance by which the ray missed its target. The halo position is also corrected for the delay of the photon, but this is a minute effect. This process is then repeated until convergence is achieved.

\section{Optical Equations}

Here we derive the equations that govern the propagation of an infinitesimal light beam through the perturbed geometry of spacetime, see \cite{Perlick:2004tq} for a textbook introduction. We work in Poisson gauge where the line element can be written as
$ds^2\!=\!a^2\! \left[-e^{2\psi} d\tau^2 - 2 B_i dx^i d\tau + \left(e^{-2\phi}\delta_{ij} + h_{ij}\right) dx^i dx^j\right]\!,$
where $\psi$, $\phi$ are the two gravitational potentials, $B_i$ is transverse and governs frame-dragging, and $h_{ij}$ is transverse and traceless and describes spin-two perturbations. In terms of $\phi$ and $\psi$ the optical equations are exact. $B_i$ is typically at least two orders of magnitude smaller than $\phi$ or $\psi$, and we therefore only keep its leading-order contribution. Its effect on the observed 
redshift is generally negligible, whereas the luminosity distance is affected at the level of $0.01 \%$ (half width at half maximum) for halos at redshift $z \sim 2$.  Our simulations also compute $h_{ij}$ but this perturbation is even smaller and we therefore choose to neglect it. The same is true for the difference between $\phi$ and $\psi$ which is of similar magnitude. Both contributions could be easily included in our ray-tracing method, but they would have no effect on our results.

The null vector $k^\mu$ tangent to the photon geodesic, and the screen vectors (the Sachs basis) $e_A^\mu$ ($A = 1, 2$) with
$g_{\mu\nu} e_A^\mu e_B^\nu = \delta_{AB}, g_{\mu\nu} e_A^\mu k^\nu = 0,$ are parallel-transported along the null ray.
For convenience we also define the complex screen vector $e^\mu = e_1^\mu + \mathrm{i} e_2^\mu$. 
The null geodesic equation
determines the path of the beam while tidal effects change the shape of the beam according to the Sachs equations
\begin{eqnarray}
 \label{eq:Sachs}
\hspace*{-0.1cm} \frac{d\theta}{d\lambda} + \theta^2 + \sigma \sigma^\ast &=& -\frac{1}{2} R_{\mu\nu} k^\mu k^\nu\,,  \quad  \frac{dD_{\!A}}{d\lambda} = \theta D_{\!A}\,,\nonumber\\
 \frac{d\sigma}{d\lambda} + 2 \theta \sigma &=& -\frac{1}{2} C_{\alpha\mu\beta\nu} e^\alpha k^\mu e^\beta k^\nu\,,
\end{eqnarray}
where $\lambda$ is the affine parameter, $R_{\mu\nu}$ and $C_{\alpha\mu\beta\nu}$ denote the Ricci and Weyl tensors, respectively, and we use the optical scalars, i.e.\ the \textit{expansion} $\theta$, the complex \textit{shear} $\sigma = \sigma_1 + \mathrm{i} \sigma_2$ and the \textit{area distance} $D_{\!A}$.

Let us introduce the direction vector $n^i$ as
\begin{equation}
 \frac{dx^i}{d\tau} = \frac{k^i}{k^0} = e^{\psi+\phi} n^i + \delta^{ij} B_j\,,
\end{equation}
which implies the normalization $\delta_{ij} n^i n^j = 1$. We also define two vectors aligned with the screen basis to form a normalized spatial triad,
$\delta_{ij} \tilde e^i_A n^j = 0$, $\delta_{ij} \tilde e^i_A \tilde e^j_B = \delta_{AB}$, with $a \, e^\mu_A = (\beta,  e^\phi \tilde e^i_A + \beta e^{\psi+\phi} n^i)$,
where $\beta$ parametrizes the freedom to choose the timelike direction to which the screen basis is orthogonal, i.e.\ all possible boosts along the direction of the ray. Since we will choose an observer at rest w.r.t.\ the CMB, $\beta$ will be of the order of a metric perturbation and can be treated on the
same footing.

In order to arrive at a convenient system of coupled ordinary differential equations (ODEs), we exploit conformal invariance by using the rescaled quantities
$\tilde{k}^0 = k^0 a^2 e^{-2\phi}$, $\tilde{D}_{\!A} = D_{\!A} e^\phi / a$, and $\tilde{\sigma} = \sigma D_{\!A}^2 e^{2\phi} / (a^2 k^0)$.
The null geodesic equation then becomes
\begin{eqnarray}
\label{eq:nullgeodesic2}
\frac{d\ln \tilde{k}^0}{d\tau} + (\psi+\phi)^\prime + 2 n^i \partial_i e^{\psi+\phi} + n^i n^j \partial_i B_j &=& 0\,,\quad\\
\frac{dn^i}{d\tau}-\left(n^i n^j-\delta^{ij}\right)\left[\partial_j e^{\psi+\phi} + n^k \partial_j B_k\right] &=& 0\,,\quad
\end{eqnarray}
and the parallel transport for $\tilde e^i_A$ is solved by
\begin{equation}
\label{eq:eparalleltransport}
\frac{d\tilde e^i_A}{d\tau} - n^i\tilde  e^j_A \left(\partial_j e^{\psi+\phi} + n^k \partial_{(j} B_{k)}\right) - \tilde e^j_A \delta^{ik} \partial_{[j}B_{k]} = 0\,.
\end{equation}
Furthermore, the Sachs equations give
\begin{multline}
\label{eq:DA}
\frac{d^2\tilde{D}_{\!A}}{d\tau^2} + \frac{d\ln\tilde{k}^0}{d\tau}\frac{d\tilde{D}_{\!A}}{d\tau} + \biggl[\frac{1}{2} n^i n^j \partial_i B^\prime_j + \frac{1}{2} n^i \Delta B_i \biggr.\\ 
\biggl. - \frac{1}{2} \left(n^i n^j - \delta^{ij}\right) e^{\phi+\psi} \partial_i \partial_j e^{\phi+\psi}\biggr] \tilde{D}_{\!A} + \frac{\tilde{\sigma} \tilde{\sigma}^\ast}{{\tilde{D}_{\!A}}^3} = 0\,,
\end{multline}
and
\begin{multline}
\label{eq:shear}
\frac{d\tilde{\sigma}}{d\tau} + \frac{d\ln \tilde{k}^0}{d\tau}\tilde{\sigma} = - \frac{\tilde{D}_{\!A}^2}{2} \left(\tilde e_1^i \tilde e_1^j - \tilde e_2^i \tilde e_2^j + \mathrm{i} \tilde e_1^i \tilde e_2^j + \mathrm{i} \tilde e_2^i \tilde e_1^j\right) \\ \times \left[e^{\phi+\psi} \partial_i \partial_j (e^{\phi+\psi} + n^kB_k) - \frac{d}{d\tau} \partial_{(i}B_{j)} \right]\,.
\end{multline}
These equations are exact in $\phi$ and $\psi$ but first order in $B_i$.
The solutions are determined by setting appropriate `final conditions' at the observer and integrating the set of ODEs backwards in time. Assuming that the observer has vanishing peculiar momentum in our coordinate system, the final conditions for the area distance and shear are
\begin{equation}
 \tilde{D}_{\!A}(\tau_o) = 0\,,\quad \left.\frac{d\tilde{D}_{\!A}}{d\tau}\right|_o = \left.-e^{\phi+\psi}\right|_o\,,\quad \tilde{\sigma}(\tau_o) = 0\,.
\end{equation}
The observables can be computed in any other inertial frame
by applying an appropriate Lorentz boost. Such a frame transformation changes the observed redshift
according to the special-relativistic Doppler effect, while the observed angles and area distances are subject to relativistic aberration. Both are straightforward transformations of the PDFs.

The final conditions for $\tilde{k}^0$ are given by an arbitrary reference frequency, e.g.\ the frequency of a spectral line that is used to determine the source redshift. For convenience we choose
$ \tilde{k}^0(\tau_o) = \left.a e^{-2\phi-\psi}\right|_o$ 
such that the observed redshift of a source becomes
\begin{equation}
 1 + z = \left. \left(\sqrt{\frac{\delta^{ij}q_i q_j}{a^2} + e^{-2\phi}} - \frac{n^i q_i}{a}\right) \frac{\tilde{k}^0}{a} e^{3\phi+\psi}\right|_s\,,
\end{equation}
where $q_i$ is the canonical peculiar momentum per unit mass for the source. Explicitly, $q_i = m^{-1}_s \partial\mathcal{L}/\partial(dx^i_s/d\tau)$, where $m_s$ is the mass of the source and $\mathcal{L}$ is the Lagrangian describing the motion of its center of mass $x^i_s$. In the non-relativistic limit $\delta^{ij} q_j / a \sim dx^i_s / d\tau$, but the above equation holds for arbitrary $q_i$.

\section{Hubble diagram}

\begin{figure}
	\includegraphics[width=\columnwidth]{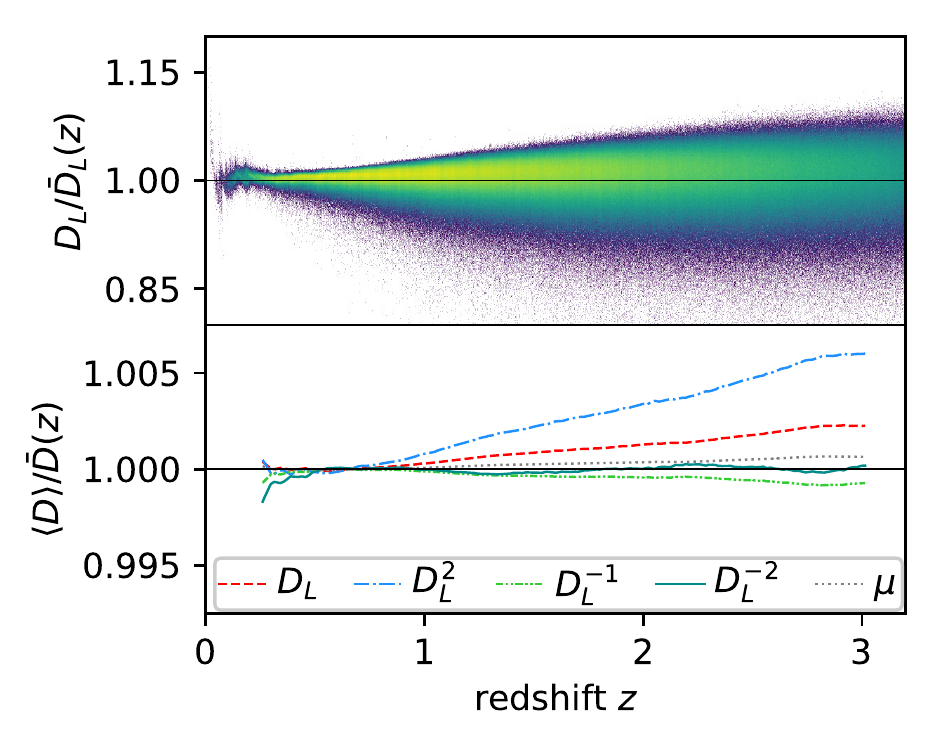}
	\caption{The top panel shows the luminosity distance $D_L$ vs.\ redshift $z$ for $\sim 10^7$ halos within a $450$ deg$^2$ field-of-view (the color gradient illustrates the logarithmic point density) in relation to the luminosity distance $\bar{D}_L(z)$ of the homogeneous $\Lambda$CDM model at the same redshift. The bottom panel shows the bias of the source average for various distance functions.}
	\label{fig:dL12}
\end{figure}

The top panel of Fig.\ \ref{fig:dL12} shows the ratio $D_L/\bar{D}_L(z)$, where $D_L = (1+z)^2 D_{\!A}$ and $\bar{D}_L$ denote the observed luminosity distance and its value in a homogeneous Friedmann--Lema\^itre (FL) model at the same observed redshift, respectively.
At low redshift the scatter is mainly due to peculiar motion, showing some indication of correlated bulk motion. On the other hand, at higher redshifts the lensing effects dominate and we have some halos that are strongly magnified while many more are slightly de-magnified.

\begin{figure}
\includegraphics[width=\columnwidth]{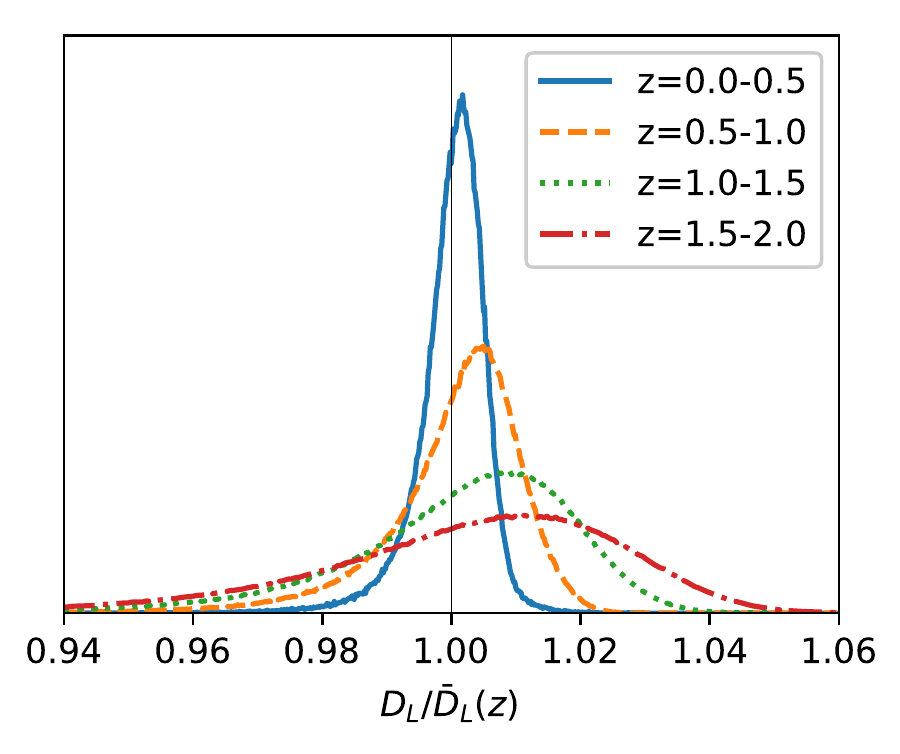}
\caption{Histograms of the ratio $D_L/\bar{D}_L(z)$ 
for four different redshift bins.}
	\label{fig:residuals}
\end{figure}

The normalized distribution of the ratio $D_L/\bar{D}_L(z)$
is shown in Fig.\ \ref{fig:residuals} for four redshift bins.
Even though the mean stays close to unity for all cases, at higher redshifts, they become increasingly non-Gaussian with a negative skew and a long tail due to the aforementioned lensing effect. We performed a resolution study on a smaller simulation volume and expect that the shapes are converged w.r.t.\ numerical effects.
This PDF has already been studied by various authors in the past, e.g.\ \cite{Wang:2002qc,Holz:2004xx,Amendola:2010ub,Macaulay:2016uwy}, but we present the first fully non-perturbative and relativistic calculation of $D_L(z)$. For $z\gtrsim 1$ the peak of the distribution is shifted by $\gtrsim 1\%$
while the shift of the mean, shown in the bottom panel of Fig.\ \ref{fig:dL12} for several choices of functions $D(D_L)$, is small as predicted from second-order perturbation theory. 
The PDFs for all choices of $D$ have a skewed shape that is important and can bias cosmological fits as we discuss next.
It is however possible to reduce the bias by binning sources in redshift~\cite{Holz:2004xx}, which ``Gaussianizes'' the likelihood.

\section{Fits to cosmological parameters}

To study the impact on cosmological parameter estimation in more detail we perform a `naive' fit to standard candles such as supernovae, using an uncorrelated Gaussian likelihood,
\begin{equation}
p(D | \theta) = \prod_{\rm{SNe}} \frac{1}{\sqrt{2 \pi \sigma_{\rm{tot}}^2}} \exp \left\{-\frac{1}{2} \frac{(\bar D - D)^2}{\sigma_{\rm{tot}}^2} \right\} \, .
\end{equation}
The product runs over all supernovae (SNe) in a sample, the distance $\bar D$ is (a function of) the usual FL luminosity distance for the values of the cosmological parameters $\theta_c = \{H_0, \Omega_m, \Omega_\Lambda\}$ to the redshift $z$ of a supernova, and $D$ is the actual distance to that supernova as obtained from our ray tracer. The total error $\sigma_{\rm tot}^2$ is composed of three contributions,
$\sigma_{\rm tot}^2 = \sigma^2_z / z + \sigma^2_0 + \sigma^2_l z$,
where $\sigma^2_z$ corresponds to a low-redshift contribution to the error from peculiar velocities, $\sigma^2_l$ models (approximately) a high-redshift contribution from lensing, and $\sigma^2_0$ is a constant `intrinsic dispersion' error. When we use $D=\mu$ (see below),  $\sigma_{\rm tot}^2$ is an absolute error, while for other choices  it is a relative error, i.e.\ $\sigma_{\rm tot}^2 \rightarrow \sigma_{\rm tot}^2D^2$. This relatively standard error model is used for example in~\cite{Betoule:2014frx} for $D=\mu$.
We estimate the $\sigma_i$ simultaneously with the cosmological parameters, i.e.\ our parameter set is $\theta = \theta_c \cup \{\sigma_z, \sigma_0, \sigma_l\}$, with a logarithmic prior for the errors and for $H_0$, and flat priors for the $\Omega_i$.

\begin{figure}
	\includegraphics[width=\columnwidth]{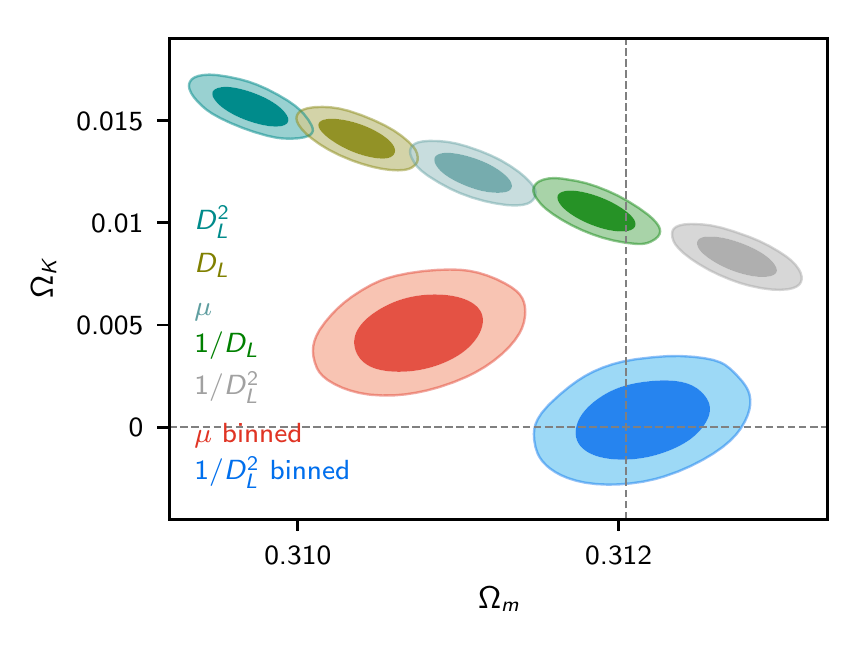}
	\caption{Constraints on $\Omega_m$ and $\Omega_K$ from fits to various functions of the luminosity distance
	(smaller contours at the top, with $D_L^2$ being the left-most contour, and then following the order of the legend until $1/D_L^2$ on the right). When binning in redshift, the bias is reduced at the prize of an increase in error bars (larger contours at the bottom for $\mu$ on the left and $1/D_L^2$ at the right -- the other functions of the luminosity distance would line up as for the smaller contours). The use of $1/D_L^2$ as distance measure combined with redshift binning is able to remove the bias from the lensing of supernovae.}
	\label{fig:omok}
\end{figure}

In order to obtain an artificial supernova-like sample, we pick $N = 5\times 10^5$ halos with uniform distribution in redshift in the range $z \in [0.023,3]$.
Usually in supernovae analysis the function $D$ is the distance modulus $\mu = 5 \log_{10} (D_L/ \mathrm{1 Mpc}) + 25$. Here we use $\mu$ and several other choices, $D = D_L$, $D = D_L^2$, $D=1/D_L$ and $D=1/D_L^2$. According to \cite{Bonvin:2015kea}, some of these are expected to be less impacted by lensing than $\mu$, but in a way that depends on the averaging performed. As we effectively average over directions for a single realization, the least affected choice should be $1/D_L^2$, see Fig.\ \ref{fig:dL12}, bottom panel.
With a Markov-Chain Monte Carlo approach we sample from the posterior of $\theta$. In Fig.\ \ref{fig:omok} we show marginalized 2d contours for $\{\Omega_m, \Omega_K = 1-\Omega_m-\Omega_\Lambda\}$.
The parameter values used in the N-body simulation are indicated by  dashed lines. The constraints that we recover depend on the choice of distance-function, but they
are biased for all choices typically at the 1\% level.
As the mean of the distances shows only a small deviation from the true background distance, cf.\ Fig.\ \ref{fig:dL12}, we expect that most of this bias is due to the non-Gaussian nature of the likelihood shown in Fig.\ \ref{fig:residuals}. It would of course be desirable to 
exploit this non-Gaussian shape since it contains information on the clustering of the dark matter, but for now we instead bin always 1000 supernovae adjacent in redshift into a single data point that, thanks to the central limit theorem, now has a much more Gaussian PDF. Indeed, as we can see from the larger, lower contours in Fig.\ \ref{fig:omok} this does reduce the bias (it should be noted that we did not 
try to optimize the binning for the trade-off of bias versus contour size). The remaining parameter shift is now due to the impact of lensing on the mean of the distance measure. But only if we combine redshift-binning with the use of $1/D_L^2$ we manage to obtain an unbiased parameter inference without having to model the lensing 
in detail.
Our full light cone only exists for a single observer position, but we performed a range of tests to verify that our conclusions do not change for other samples: We selected different halos for our supernova-like sample, creating effectively new data sets, and we also started at higher redshifts, $z_{\rm min} = 0.1, 0.2$ and $0.3$, instead of the value of $z_{\rm min} = 0.023$ commonly used, to test for dependence on the local observer environment.

\section{Conclusions}

For the first time we provide a non-perturbative and fully relativistic numerical calculation of the observed luminosity distance and redshift for a realistic cosmological source catalog in  $\Lambda$CDM cosmology. The catalog is comprised of dark matter halos from a large, high-resolution N-body simulation that fully captures the general relativistic dynamics and provides an accurate description of the metric on the past light cone of an observer.
The observables are computed by integrating the optical equations along the true photon trajectories obtained by a shooting method. While we work in a weak-field context that allows us to neglect certain terms, at no point do we take a Newtonian limit. We solve the scalar part of the geodesic and optical equations exactly while we keep the frame-dragging terms only to leading order and neglect gravitational waves altogether. 

Our numerical experiment provides conclusive evidence that the relativistic evolution of inhomogeneities, once consistently combined with the kinematics of light propagation on the inhomogeneous spacetime geometry, does not lead to an unexpectedly large bias of the distance-redshift correlation.
This corroborates the conclusions of \cite{Clarkson:2011br,Kaiser:2015iia,Fleury:2016fda,Kibble:2004tm,BenDayan:2012wi,Clarkson:2014pda,Adamek:2017mzb}. On the other hand, inhomogeneities introduce a significant non-Gaussian scatter that can give a large standard error on the mean when only a small sample of sources is available. But even for large, high-quality samples this scatter can bias the inferred cosmological parameters at the per-cent 
level. This can only be avoided if the shape of the distribution is properly characterized and included in the analysis, or by binning the supernovae in redshift to obtain a more Gaussian scatter. In either case we should also use $1/D_L^2$ as distance measure to minimize the bias in the mean distance.
 Finally, we want to stress that our approach is conceptually much more powerful than previous ones because it can be consistently extended beyond the $\Lambda$CDM concordance model, including settings that have no suitable Newtonian or quasi-static limit \cite{Sawicki:2015zya}. An exploration of this potential is left to future work.

\bigskip

\begin{acknowledgments}
We thank David Daverio for discussions and Peter Behroozi for useful advice concerning the light-cone feature of \textit{ROCKSTAR}.
This work was supported by a grant from the Swiss National Supercomputing Centre (CSCS) under project ID s710. Some post-processing was carried out on the \textit{Baobab} cluster at the University of Geneva. JA and CC are supported by STFC Consolidated Grant ST/P000592/1. RD and MK acknowledge financial support by the Swiss National Science Foundation. We acknowledge the use of \textsc{getdist} (originally part of \textsc{CosmoMC} \cite{Lewis:2002ah}) for some of the plots.
\end{acknowledgments}

\bibliography{hubble}

\end{document}